\documentclass[preprint,1p,times,floatfix]{elsarticle}
\RequirePackage{lineno}
\usepackage{graphicx}
\usepackage{amssymb}
\usepackage{wrapfig}
\usepackage{floatflt}
\usepackage{setspace}
\def\deg{^\circ}

\begin{document}

\title{Fundamental Neutron Physics Beamline at the Spallation Neutron Source at ORNL}

\author{N.~Fomin$^{1}$}
\author{G.~L.~Greene$^{1,2}$}
\author{R.~Allen$^{2}$}
\author{V.~Cianciolo$^{2}$}
\author{C.~Crawford$^{3}$}
\author{T.~Ito$^{7}$}
\author{P.~R.~Huffman$^{2,4}$}
\author{E.~B.~Iverson$^{2}$}
\author{R.~Mahurin$^{5}$}
\author{W.~M.~Snow$^{6}$}
\address{$^{1}$University of Tennessee, Knoxville, TN, USA}
\address{$^{2}$Oak Ridge National Laboratory, Oak Ridge, TN, USA}
\address{$^{3}$University of Kentucky, Lexington, KY, USA}
\address{$^{4}$North Carolina State University, Raleigh, NC, USA}
\address{$^{5}$University of Manitoba, Winnipeg, Manitoba, Canada}
\address{$^{6}$Indiana University and Center for the Exploration of Energy and Matter, Bloomington, IN, USA}
\address{$^{7}$Los Alamos National Laboratory, Los Alamos, NM, USA}

\begin{abstract}

We describe the Fundamental Neutron Physics Beamline (FnPB) facility located at the Spallation Neutron Source at Oak Ridge National Laboratory.  The FnPB was designed for the conduct of experiments that investigate scientific issues in nuclear physics, particle physics, and astrophysics and cosmology using a pulsed slow neutron beam. We present a detailed description of the design philosophy, beamline components, and measured fluxes of the polychromatic and monochromatic beams.

\end{abstract}
\maketitle

\section{Introduction}

Cold neutrons and ultracold neutrons (UCN) have been employed in a wide variety of experiments that shed light on important issues in nuclear, particle, and astrophysics. These include the determination of fundamental constants, the study of fundamental symmetry violation, searches for new interactions in nature and tests of the fundamental laws of quantum mechanics. Their special combination of properties make them a good choice to address the expanded list of fundamental scientific mysteries which now confront us in the wake of the discoveries of dark matter and dark energy, which shows that 95\% of the energy content of the universe resides in some unknown form.  In many cases, experiments with slow neutrons provide information not available from existing accelerator-based nuclear physics facilities or high-energy accelerators~\cite{Nico05b, Abele:2008, Dubbers:2011}. 

For this reason, most major neutron user facilities have included, as a component of their scientific program, the investigation of fundamental interactions. The great majority of these facilities have been located at continuous-wave (CW) spallation sources or intense research reactors to take advantage of the highest neutron intensities available. However, many precision experiments could benefit from the special characteristics of pulsed mode spallation neutron sources, which enables experiments to use the time and energy structure of the slow neutron beam to advantage. The peak and average cold neutron fluxes from pulsed reactors like the Frank Laboratory of Neutron Physics and spallation sources like the Los Alamos Neutron Science Center (LANSCE), the Spallation Neutron Source (SNS), the Japanese Spallation Neutron Source (JSNS), the ISIS spallation source at the Rutherford Appleton Lab, and the future European Spallation Source (ESS) are now high enough that a relatively broad class of fundamental neutron physics experiments can be best performed at such facilities.

In this paper we describe the construction and commissioning of a pulsed slow neutron beamline and facility at the Spallation Neutron Source (SNS) at Oak Ridge National Laboratory which is optimized to conduct certain experiments in this field. The design of the facility attempted to preserve the advantages of pulsed spallation sources in reducing systematic errors in a broad class of fundamental neutron physics experiments. One such feature is the well-known time structure of the beam.  It can be used to identify and analyze the neutron energy dependence of background signals which are invisible at a CW source, to shape the beam phase space in special ways using time-dependent neutron optical components, or to help perform absolute neutron polarization measurements.  Another feature of a spallation neutron source is the fact that the neutron source is off by the time the slow neutrons arrive at the apparatus, which generally reduces background signals in detectors. With improved neutron optics technology one can bend the slow neutron beam enough to eliminate line-of-sight to the production target and to guide it far from the spallation source so that the very high energy neutrons and gamma rays generated in spallation do not adversely affect the physics experiments. 

Furthermore, since this beamline is operated as a user facility with all beam time allocated on the basis of independent peer reviews, it was essential to design the facility to accommodate a wide variety of different types of slow neutron experiments and especially to avoid precluding to the extent possible new unforeseen ideas for future experiments. An Instrument Development Team (IDT) for the beamline facility reached a consensus based on experience at other facilities combined with feedback and discussions from potential future users in the form of the following design principles as follows:
\begin{enumerate}
\item Total intensity was to be maximized. Essentially all of the experiments in this field are limited by statistics. The facility design must not include features that lead to a permanent reduction in the neutron fluence (total number per second). 

\item Cold neutron intensity is of the highest priority. Almost all of the experiments of relevance for the areas of nuclear and particle physics prefer low energy neutrons. Slower neutron beams raise the signal/background ratio in neutron decay experiments. The opportunities for creative manipulation of the beam properties (phase space, polarization, etc.) are greater for cold neutrons. 

\item The unique properties of a spallation neutron source should not be compromised in the beamline design. The advantages that a spallation source offers experiments in this area relative to a steady-state source are directly or indirectly rooted in the built-in use of neutron time-of-flight and a corresponding potential for increase in the signal/background ratio. Everything must be done to preserve this advantage while at the same time being consistent with (1) and (2).

\item Accommodate the different demands of different classes of experiments consistent with (1)-(3).

\item Leave as much floor space as possible.  This allows for flexibility in the design of future experiments.

\end{enumerate}
In the remainder of this paper we describe the details of the design of the facility which we have realized consistent with these principles and the physical constraints and properties specific to the Spallation Neutron Source at ORNL. We hope that this detailed description of the beamline design and properties will be useful both to potential scientific users and to the design of possible future facilities at pulsed spallation sources such as the ESS~\cite{Snow:2014}.
 The specific design choices made for the SNS facility are by no means unique: we encourage the reader to contrast the SNS facility with the other fundamental neutron beamlines constructed at pulsed spallation neutron source, namely the pioneering beamline at LANSCE~\cite{Seo04} and the fundamental neutron physics beamline at the JSNS, which supplies a collection of three slow neutron beamlines with different phase space profiles optimized for different subclasses of experiments~\cite{Arimoto:2012zma, mishima09}. We also encourage a comparison with the intense CW slow neutron beam facilities for fundamental neutron physics already in operation or under development at PSI~\cite{Fischer97, Bauer98, Schebetov03, Zejma05}, ILL~\cite{Abele06}, and NIST~\cite{Nico05a, Cook09}.


 


%
%
%

\section{The Spallation Neutron Source}

The Spallation Neutron Source (SNS) is the most intense pulsed neutron
source in the world, designed to deliver a sub-microsecond proton
pulse onto a mercury target at a repetition rate of 60~Hz with
time-averaged proton power of 1.4~MW.  The released spallation
neutrons are moderated by supercritical hydrogen and water moderators.  The resulting
slow neutrons are used for a variety of experiments.  The SNS supports 24 instruments which can conduct experiments simultaneously.

The SNS uses a cesium-enhanced, RF-driven multi-cusp ion source~\cite{stockli2010} to provide a 65~keV H$^{-}$ beam at 60~Hz with a pulse length of up to 1~ms.   The normal conducting linac consists of six drift tube linac (DTL)
tanks, which bring the H$^{-}$ beam energy up to 86.8~MeV, and four coupled
cavity linac (CCL) structures, which supply additional acceleration to
bring the energy to 185.6~MeV.  This is followed by a superconducting
linac consisting of 11~medium-beta ($\beta$=0.61) cryomodules and
12~high-beta ($\beta$=0.81) cryomodules.  The medium-beta cryomodules
each contain 3~cavities and provide 10.1~MV/m, bringing the H$^{-}$ ions
upto an energy of 379 MeV.  The high-beta cryomodules (consisting of
4~cavities each) are designed to provide upto 15.9 MV/m resulting in a maximum proton energy of 1.3 GeV.  In June of 2014, the SNS was operating at 940~MeV of proton energy.

The H$^{-}$ pulses arrive at the ring injection point via a 150-m long
beam line which is used for energy collimation (bending magnets) and
transverse halo collimation (straight sections).  The ring consists of
four straight sections as well as four arcs, with a total flight path corresponding to 1 us of accumulated proton pulse length.  The incoming H$^-$ beam
is stripped of its two electrons by passing through a thin carbon or
diamond foil, and is merged in phase space with the circulating proton
beam.  Currently, the ring design allows operation at up to 1.0~GeV, but RF
systems and injection kickers have been designed to support 1.3~GeV
operation with minimal upgrades.

\begin{figure}[htb] 
\begin{center}
  \includegraphics[width=.6\textwidth]{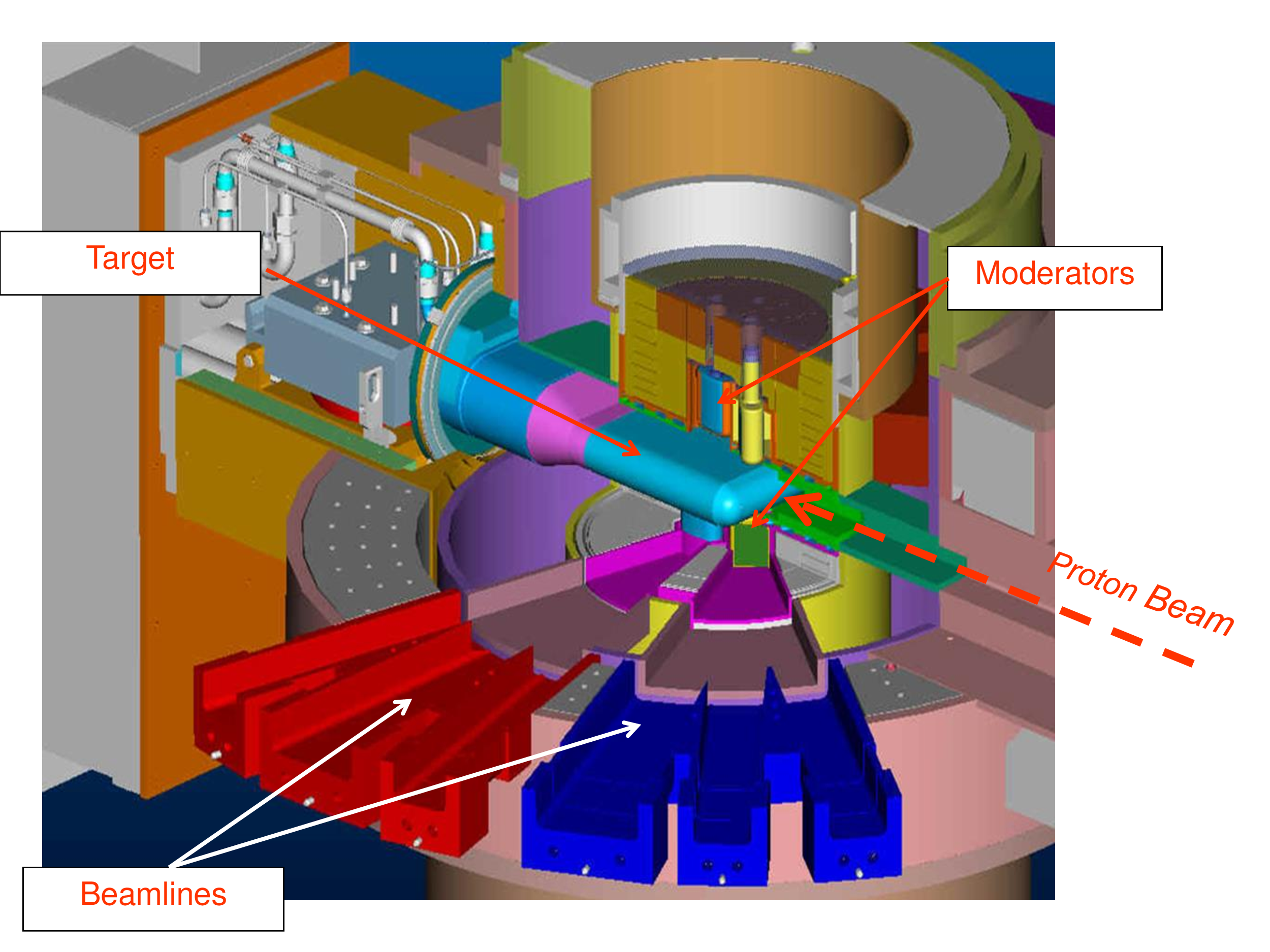}
\end{center}
  \caption{A cutaway view of the mercury target, surrounded by four
    moderators enclosed in a beryllium reflector.  Each moderator is
    viewed by several beamlines.  See text for details.}
\label{target_big}
\end{figure}

\begin{figure}[htb] 
  \includegraphics[width=.9\textwidth]{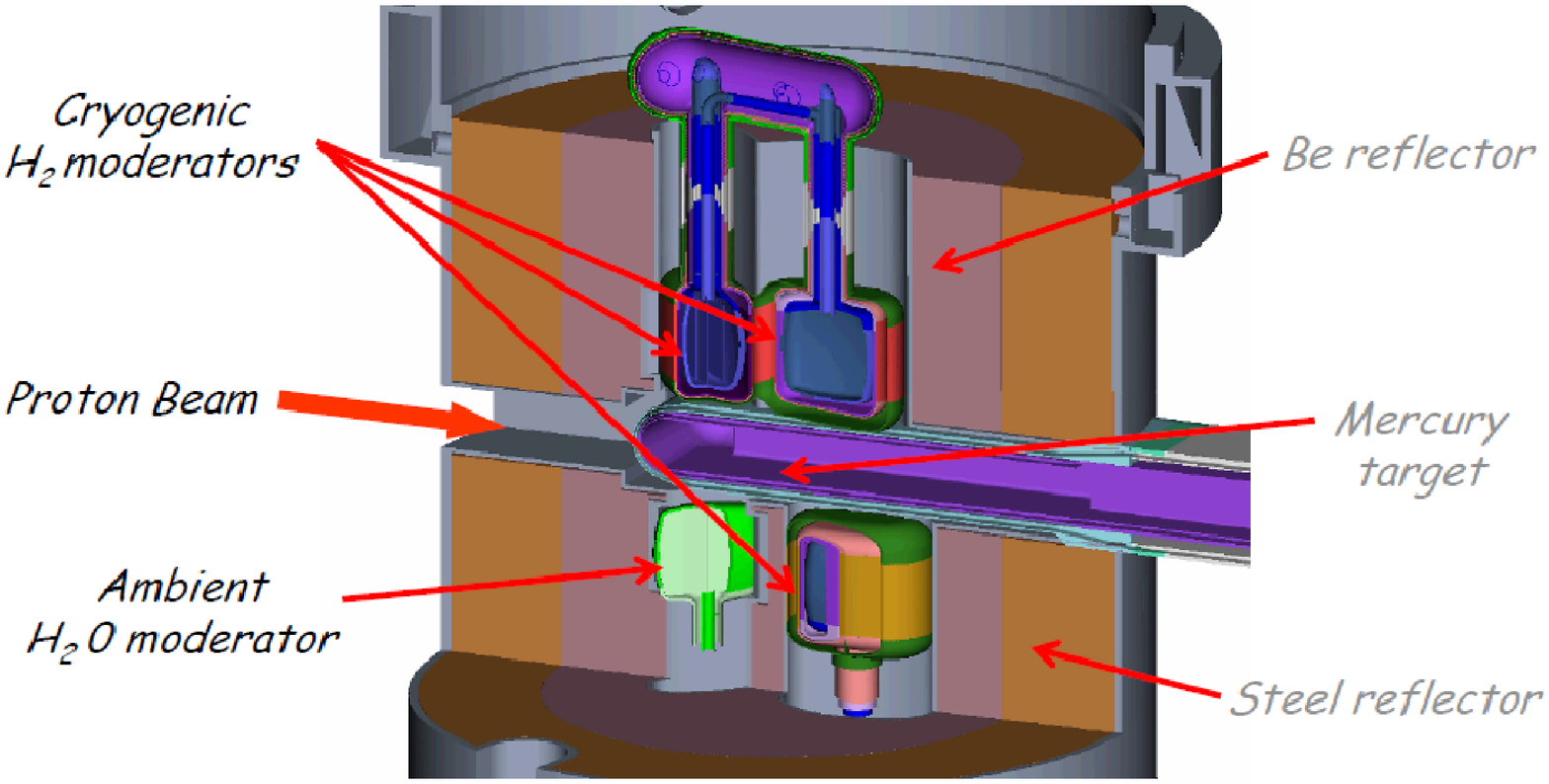}
  \caption{The mercury target vessel surrounded by 4 moderators inside
    of a beryllium reflector.  FnPB views the bottom downstream
    moderator (bottom right in the figure). See text for details.}

\label{reflector_moderators}
\end{figure}

The protons are incident on a target of circulating mercury, producing spallation reactions as they deposit their energy.
The mercury is contained inside the stainless steel target module, one of the components of the target system.  The target module is made of two concentric vessels, where the inner vessel contains the target mercury during normal operation, and the outer vessel contains any mercury that may leak from the inner vessel.  The main process loop contains $\approx$1.4~m$^3$ of mercury, circulating at $\approx$325 kg/s, at a pressure of $\approx$0.3 MPa. The inner vessel is cooled by flowing mercury, whereas the outer vessel is cooled by water.  The two vessels are separated by a helium-filled interstitial region, where there are two instruments present.  A heated resistance temperature detector (RTD) is able to detect leakage as well as distinguish between mercury and water; an electrical conductivity probe detects the presence of mercury between the contacts~\cite{completion_report,Henderson2014}.  
 
Spallation neutrons are moderated by undergoing repeated scattering primarily with hydrogen and beryllium atoms.  The configuration at
the SNS includes four moderators, two of which are viewed from both
sides of the target.  Three moderators are supercritical hydrogen at $\approx$ 20 K, and one is liquid water at $\approx$ 320 K.  The moderators are surrounded by a
water-cooled beryllium inner reflector, which is then surrounded by water-cooled
stainless steel~\cite{completion_report,sns-sct-base02}.  The
configuration can be seen in Fig.~\ref{reflector_moderators}.  The
viewed faces of all the moderators are 10 cm (horizontal) by 12 cm
(vertical).

 The FnPB views the bottom downstream moderator, where the hydrogen is
 delivered through the bottom of the vessel via a jet, which forces
 the hydrogen to circulate.  This moderator is fully coupled
 unpoisoned hydrogen (nominally parahydrogen) at 20 K (viewed from one side
 only), with a curved viewing surface  and maximum moderator thickness
 of 60 mm (average of 55 mm).  The moderator is surrounded by
 $\approx$20 mm of light water acting as a premoderator.

\section{Fundamental Neutron Physics Beamline - Overview}
\begin{figure}[h!] 
  \includegraphics[width=.8\textwidth]{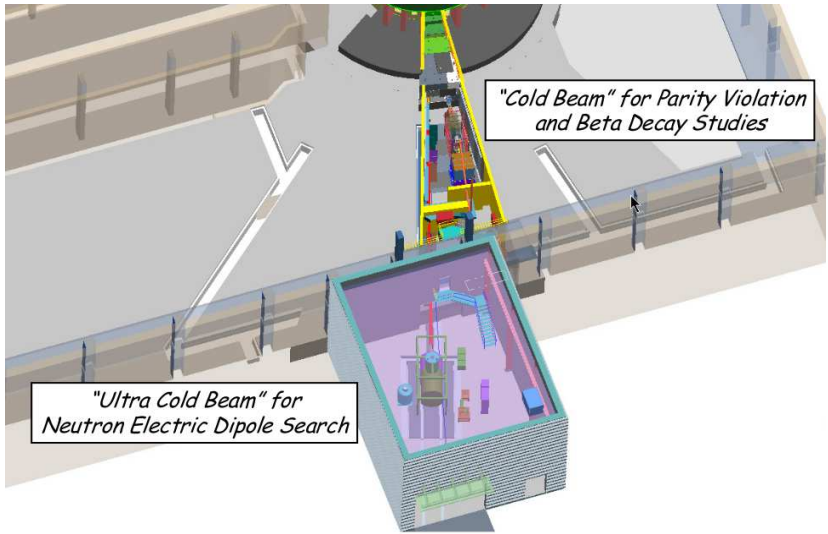}
  \caption{A top view of the Fundamental neutron Physics Beamline(s).
    13A is the monochromatic line (left) and will extend into the
    external building to serve the nEDM experiment. 13B is the
    polychromatic beamline (right), shown with the NPDGamma experiment ~\cite{npdg_proposal,snow00,Fomin:2013ssa} installed in
    the experimental area.}

\label{top_view_bl_3d}
\end{figure}
\begin{figure}[h!] 
  \includegraphics[width=.8\textwidth]{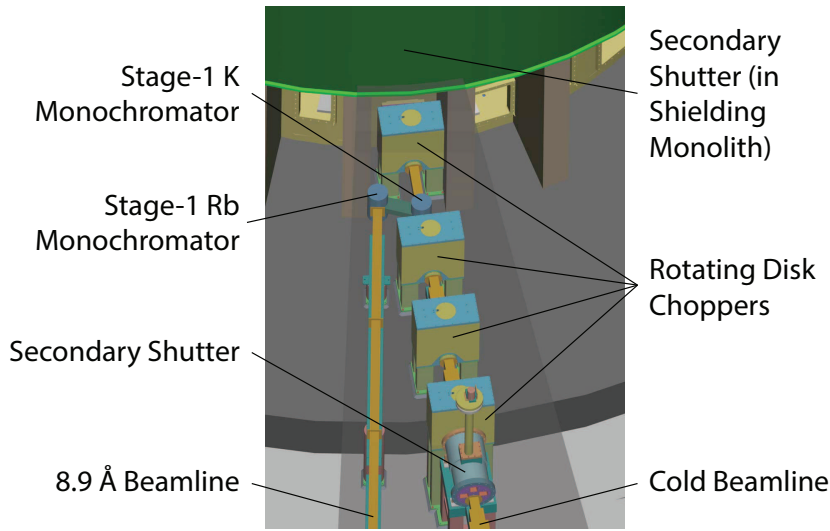}
  \caption{Schematic view of the Beamline 13 components, including
    bender guide, choppers, monochromator assembly, secondary shutters
    (cold and UCN lines), as well as straight (A/cold) and expanding (B/UCN) guides. See text for details.}

\label{top_view_bl}
\end{figure}

The Fundamental Neutron Physics Beamline (FnPB) is one of three experimental areas which view this coupled liquid hydrogen moderator. These areas are ``pie-like'' slices, surrounding the SNS target monolith.  An overhead view of the FnPB is presented in Fig.~\ref{top_view_bl_3d}. FnPB delivers neutrons
to two smaller experimental areas via a beamline which is split shortly after it exits the target shielding monolith.  Beamline 13A, or the ``UCN'' line, delivers 8.9~$\AA$ neutrons to an external building expected to host an experiment to search for the neutron electric dipole moment~\cite{Lamoreaux:1995pq}.  These neutrons will be converted to UCN via the super-thermal processes in $^4$He~\cite{golub1975}. The external FnPB building begins at the outside wall of the SNS target building.  Its volume is 2789.2~m$^3$ with a useful floor area of 220.18~m$^2$ and height of the crane hook 7.931~m.  The bridge of the crane is parallel to the wall of the SNS target building. The nEDM experiment~\cite{Lamoreaux:1995pq} requires an isolation pad, which is made of 1.22~m thick concrete, is 4.62~m wide and 7.28~m long, begins 21.90~m from the end of the UCN guide and is oriented so that its center is along the line of sight to the UCN guide. There is a second isolation pad to support the neutron guide (not yet installed) that begins at the external building boundary and extends to the main isolation pad.  The guide isolation pad is made of 0.61~m thick concrete and is 1.30~m wide and 4.44~m long.  The boundaries between the isolation pads and the external building floor are filled in with styrofoam.

 Beamline 13B, or the ``cold'' beamline, is polychromatic and completely enclosed in a cave in the SNS target
building. The broad distribution of slow neutron energies in this beam can serve a suite of experiments.  The boundaries of the experimental area are shown in yellow in Fig.~\ref{top_view_bl_3d}, with two labyrinth walls defining the entrance. The useful space is 2.84~m across on the upstream end of the cave and 5.38~m on the upstream end of the first labyrinth wall, with 12.65~m along the direction of the beam.  The experimental area has a pit in the floor to accommodate a large magnetic spectrometer for the Nab experiment~\cite{nab_proposal,dinko09}.  The pit is 4.88~m long, 2.13~m wide, 2.44~m deep and begins 16.02~m downstream of the moderator face.


%
%
%
%
%
\subsection{Shielding}

An extensive shielding package that satisfies the SNS radiological requirement of $<$0.25~mrem/hr combined
gamma and neutron radiation at the boundary between adjacent beamlines
and in the instrument hall at 2~MW power extends from the moderator outwards to the first labyrinth wall.  The individual shield blocks were designed so that there is no direct line of
sight through joints extending from the source to the experimental area.
The shielding configuration starts with 2.5-5.1~cm of steel around the
guide, which staggers periodically along the beamline to prevent neutrons and gammas streaming through gaps in the shielding.  The steel is followed by
high density (HD) concrete up to 9 m from the moderator.  The height of the beamline shielding is
4.45~m from the instrument floor and extends from the source out to a distance of 7.4~meters.  Here, the shielding height drops to 3.96~m  and continues to the experimental cave, $\approx$15.2~m downstream of the moderator.  The cave is covered by a 45~cm thick regular concrete roof. A 4.5~m$^3$ heavy concrete cube beamstop begins at 6.91~m from the upstream end of the cave. Its support stand is 1.78~m wide, 1.65~m deep, 96.5~m tall and made of regular concrete.  On the upstream face of the concrete beamstop, a 0.30$\times$0.3$\times$0.3~m$^3$ cutout has been made to accommodate a lithium carbonate neutron beam stop at the rear.  This cutout is made at beam height, meaning its center is 1.8~m from the cave floor.

In addition to the radiological shielding requirements, SNS
guidelines also dictate that the magnetic field should be $<$50 mG at
the boundaries between adjacent beamlines.  The FnPB itself requires no modifications to comply with this requirement.  However, before initial commissioning of 13B, magnetic shielding was installed for the NPDGamma experiment~\cite{npdg_proposal,snow00,Fomin:2013ssa}.  The walls and ceiling of
the experimental cave are lined with 0.635~cm plates of A1010 steel, whereas the walls are lined with multiple plates to give 2.54~cm on the beam-left side and 2.54-5~cm on the beam-right side.



\subsection{Neutron Guide}
%
%



Slow neutrons can be reflected from polished surfaces of various materials. For most materials,  which possess neutron refractive indices slightly smaller than 1, neutrons will undergo total external reflection from the optical potential $V_{opt}$ of the medium if their angle of incidence is below a critical angle $\theta _c$, determined by the material density, the neutron coherent scattering length, and the neutron wavelength.  This critical angle is set by the condition $q^{2}/ 2m < V_{opt}$ where $q$ is the momentum transfer perpendicular to the surface.  In the case of natural nickel, the critical angle is $\theta _c =1.73 \mbox{mrad}*\lambda$, where $\lambda$ is given in $\AA$.  The critical momentum transfer is then $q_{c,Ni}$=$\frac{4\pi}{\lambda}\sin\theta _c$=0.0217~$\AA^{-1}$.  Supermirrors are multilayers of reflective materials arranged to increase the reflectivity above the critical angle of a uniform medium through interference scattering. The effective critical angle for a supermirror is commonly specified by a parameter $m$ given in multiples of $q_{c,Ni}$.  

The upstream portion of the neutron guide is curved in order to minimize background from fast neutrons and gammas inside the experimental area.  The beamline begins 1~m from the face of the hydrogen moderator with
a straight, rectangular shaped "core" guide that is 10~cm in width and 12~cm in height.  This first guide element is a
1.275~m long, $m = 3.6$ supermirror guide.

Following the core guide is a $\sim 4.5$~m long, five channel bender with a 117~m radius bend towards the beam-left
direction.  The bender is composed of a series of straight, $\sim 0.5$~m long sections of guide, each rotated about
0.22$^{\circ}$ relative to the previous section.  Each segment consists of 5 vertical channels, separated by
partitioning septa of thickness 0.55~mm.  The top, bottom, and beam-right faces of the guide are coated with $m=3.8$
supermirror.  The beam-left faces are coated with $m = 2.3$ supermirror.  The first 1.8~m of this bender (four sections) is located in the primary shutter housing and is translated out of the beam path when the shutter is closed.  The next
1.2~m of bender guide contains three sections and extends from the shutter housing to the first chopper housing.  An
additional 1.2~m (three sections) of bender follows from the chopper to the monochromator.  Finally, 0.3~m of bender
extends from the monochromator housing to the straight guide.  Loss of direct line-of-sight to the moderator occurs at 
a distance of 7.5~m from the moderator.

The remaining guide is straight and extends from the end of the bender guide to the experimental area, terminating at
15~m from the face of the moderator.  There are small gaps in this guide for the remaining three choppers as well as the
secondary shutter.  All surfaces of the straight guide are $m=3.6$ supermirror and there are no partitioning septa.

All of the guide sections are enclosed in a vacuum tube evacuated with a dry roughing pump. All of the guide substrates are glass except the core guide, which is polished aluminum cooled with helium gas to lower the temperature and suppress possible interdiffusion of the supermirror layers which could degrade their reflectivity.


\subsection{Choppers}
 The neutron choppers are rotating disks (axis of rotation parallel to the neutron beam) coated with a neutron-absorbent material, containing one or more apertures to allow neutrons to pass through. By adjusting the chopper phase relative to the proton pulse such that the aperture is aligned with the beam when neutrons of a desired energy (corresponding to a particular time of flight to the chopper) arrive at the chopper -- a neutron energy band can be selected. 

The FnPB includes two choppers located 5.5 m and 7.5 m downstream of the
moderator, as well as housing to accommodate two more choppers at 9 m
and 10.5 m if needed for specific experiments.  The gaps in the guide are 58 mm for the first chopper position, and 56 mm for the remaining 3 positions. The FnPB choppers spin at 3600 RPM as
the neutron pulses are delivered at 60 Hz. The chopper disks are made of
carbon fiber composite with a coating containing $^{10}$B (minimum of
0.13 g/cm$^2$) to absorb neutrons. The physical parameters of the
choppers are listed in Table.~\ref{chopper_table}. 

\begin{table}[htpt]
\begin{center}
\caption{Parameters of the FnPB choppers}
\vspace*{0.25in}

\begin{tabular}{|c|c|c|}
\hline
 & Chopper 1& Chopper 2 \\
\hline
Axis to beam center & 25.0 cm & 25.0 cm\\
Outer diameter & 63.7 cm & 63.7 cm\\
Cutout Angle & 131$\deg$ & 167$\deg$\\
Cutout inner radius & 18.6 cm & 18.6 cm \\
\hline
\end{tabular}
\label{chopper_table}
\end{center}
\end{table}

\begin{figure}[h!] 
\begin{center}
  \includegraphics[angle=270,width=.8\textwidth]{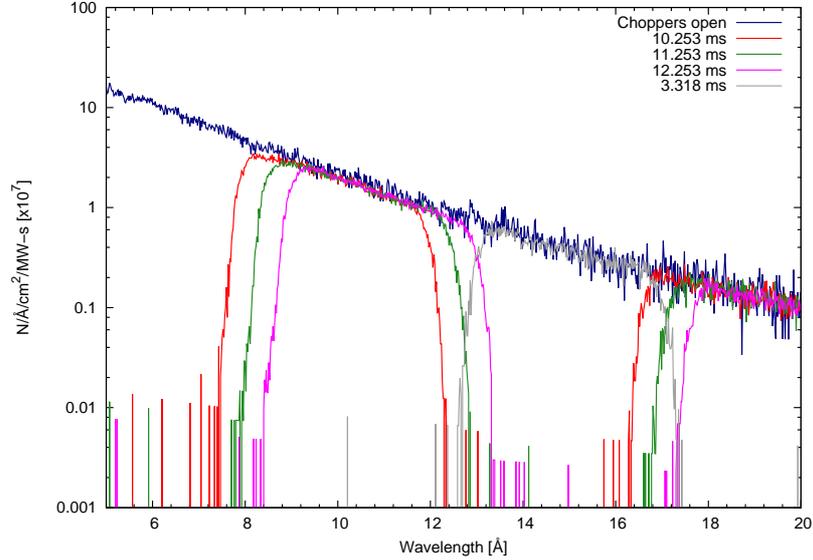}
\end{center}	
  \caption{Unchopped spectrum (both choppers parked open) is shown in blue.  The other spectra correspond to chopper 1 parked open and chopper 2 running with the phase delays indicated. }

\label{fig:chopped}
\end{figure}

An example of a chopped spectrum is shown in Fig.~\ref{fig:chopped}.  Here, the first chopper was parked open, and the phase delay of the second chopper relative to the proton pulse was scanned.  These data were taken during a special 5~Hz operating mode of the SNS, which is normally operated at 60~Hz.

\subsection{Monochromator Crystals}
A double-crystal monochromator system based on the principles outlined in Ref.~\cite{Mattoni2004343} was designed to direct the 8.9~$\AA$ beam towards the external building.  If the two reflections in the double-crystal monochromator were from lattices with the same $d$-spacing, the outgoing monochromated beam would emerge parallel to the cold beamline at its intersection with the first monochromating crystal. Given the location of the external building and the desire to maximize the floor space for cold neutron experiments, one would ideally like for the monochromatic beam to be directed along the side wall of the 13B enclosure, which is at an angle of $\sim$10$^{\circ}$ relative to the cold beamline.  By using a combination of two different intercalant atoms~\cite{Courtois2011S37}, and thus crystal spacings, we direct the monochromated beam at an angle of 9$^{\circ}$ relative to the cold beamline, along the said side wall, maximizing the separation of the two beamlines.


The first monochromator is stage-1 potassium intercalated graphite, and
consists of an array of 24 crystals, each having dimensions 20 mm x 45
mm for a total area of 120 mm x 180 mm.  It intersects the full beam
(100 mm x 120 mm) and reflects neutrons of 8.90~$\AA$ with high probability, as well as
$\lambda/n$ wavelengths with lower probability.  A mosaic of $\sim$3$^{\circ}$ was chosen to match the divergence of the neutron guide.

\begin{figure}[h!] 
\begin{center}
  \includegraphics[width=.85\textwidth]{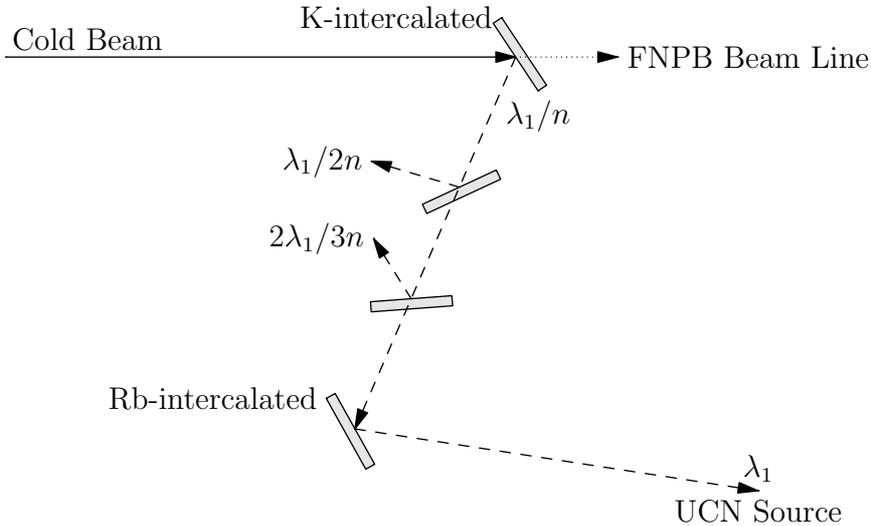}
\end{center}
\caption{Top view of the monochromator.  First monochromator crystals are potassium intercalated in graphite, second monochromator crystals are rubidium intercalated in graphite, with pyrolytic graphite crystals in between.  See text for details.}
\label{mono_diagram}
\end{figure}

Between the two monochromators, pyrolytic graphite crystals were inserted to remove the unwanted $\lambda/n$ wavelength neutrons,  one oriented to reflect $\approx$4.45$\AA$ ($\lambda/2$) neutrons and the
other oriented to reflect 2.97$\AA$ ($\lambda/3$) neutrons. The mosaic widths (5$^{\circ}$) of these
crystals are broad enough that precise wavelength tuning is not
required. Note that neutrons of 2.23$\AA$ ($\lambda/4$)
are also reflected in second order by the $\lambda/2$ filter.

The second monochromator, stage-1 rubidium intercalated graphite, consists
of an array of 35 crystals, each having dimensions 20~mm~x~45~mm for a
total area of 140~mm~x~225~mm.  This monochromator reflects the beam
of 8.90$\AA$ neutrons (as well as any remaining $\lambda/n$ neutrons)
into the UCN ballistic guide.  The orientations of the crystals as well as the primary and reflected beams are shown in Fig.~\ref{mono_diagram}.  Detailed characterization measurements were carried out at the FRM2 reactor in Munich with results presented in Ref.~\cite{Courtois2011S37}.

\begin{figure}[h!] 
\begin{center}
  \includegraphics[width=.85\textwidth]{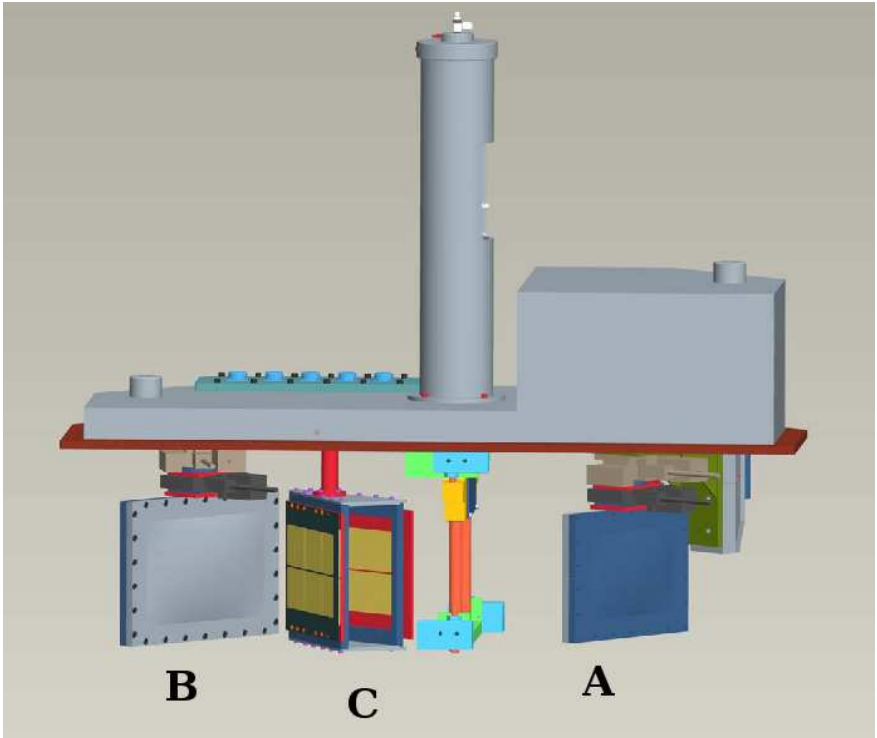}
\end{center}
\caption{Monochromator crystal assemblies: (A) potassium intercalated
  in graphite, (B) rubidium intercalated in graphite, and (C) graphite
  $\lambda/n$ filters, along with their housing.  On the right, crystal assembly A
  can be retracted into the top of the housing when BL13A is not
  taking data. See text for details.}
\label{mono_inside}
\end{figure}

The monochromator housing begins 6.5~m downstream of the moderator and is accommodated by a 29.6~cm gap in the guide.
Fig.~\ref{mono_inside} shows the arrangement of the crystals inside
the monochromator housing.  All components are mounted to the top flange of the chamber for ease of service and in addition, the chamber itself is designed to allow for
the first crystal assembly to be retracted from the cold beam when not in use to avoid neutron damage as well as neutron loss in the cold beam.

\subsection{Cold Beamline}

%
%
%
%
A secondary shutter on the cold beamline is located beyond the
end of the loss of sight to the moderator, starting 10.5 m downstream of the moderator face. It consists of a steel drum with a rotating cylinder inside containing 0.5 m of
neutron guide in the open position and 0.48 m of steel in the closed position. The upstream end of the steel includes a a beamstop of 2~cm think $^6$Li-phosphate tile weighing 218~g and containing
33.13~g of $^6$Li.

\begin{figure}[h!] 
  \includegraphics[angle=270,width=.8\textwidth]{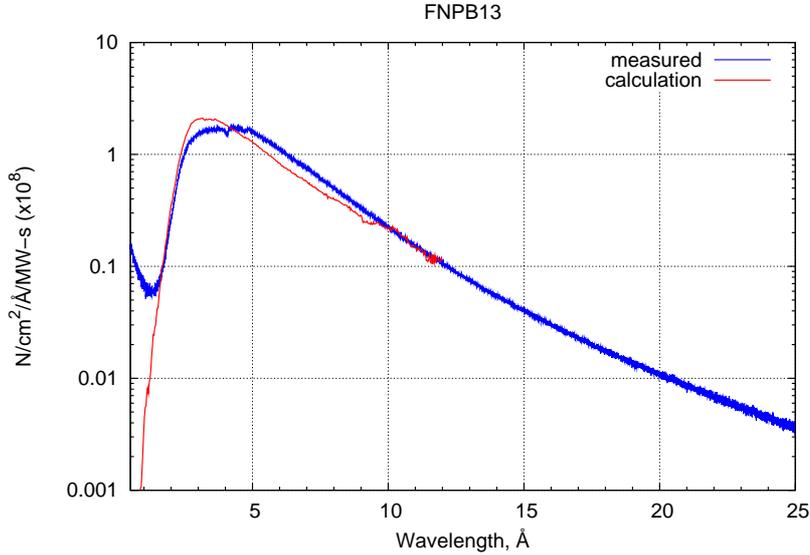}
\caption{Measured and calculated neutron brightness per MW of proton power on the spallation target for the cold guide at the FnPB as a function of neutron wavelength.}
\label{cold_spectrum}
\end{figure}

The spectral flux was measured at the end of the cold beamline
15.15~m from the moderator face using  an efficiency calibrated $^3$He proportional counter~\cite{ip:icansxvii-iverson-1}.   As the neutrons make several bounces off the walls of the guide, the beam phase space is presumed to have little position dependence over the area of
the guide exit.  The energy of the neutrons was calculated based on
time of flight from the moderator to the $^3$He detector and
calibrated using aluminum Bragg edges from windows
along the beamline.  This measured spectrum is compared to the flux
spectrum calculated using a McStas model of the
beamline~\cite{huffman2005} in Figure~\ref{cold_spectrum}.

\subsection{UCN beamline}
The UCN beamline continues downstream of the monochromator assembly
with an 8 m section of ``ballistic'' guide~\cite{tito06} in order to optimize neutron transport over long distances.  The guide starts with a rectangular
cross section 12~cm wide and 14~cm tall and expands to 20~cm by 30~cm.  The top and bottom of the guide have reflectivity of $m=3.6$ and the sides have $m=2.2$.

The transverse distribution of neutrons passing through the
monochromator assembly was measured at the end of the 8~m long
ballistic section of the UCN guide.  Several techniques were employed,
including using the previously mentioned absolutely calibrated $^3$He
detector in combination with neutron-sensitive image plates, as well
as a CCD camera~\cite{ip-hamilton-2012}.  The time-of-flight spectrum was measured with a $^3$He detector behind a 1.27 cm radius pinhole, to keep the rates down.  The image plates and CCD camera were used to establish and confirm the fraction of the total flux seen through the pinhole.

\begin{figure}[h!] 
  \includegraphics[angle=270,width=.8\textwidth]{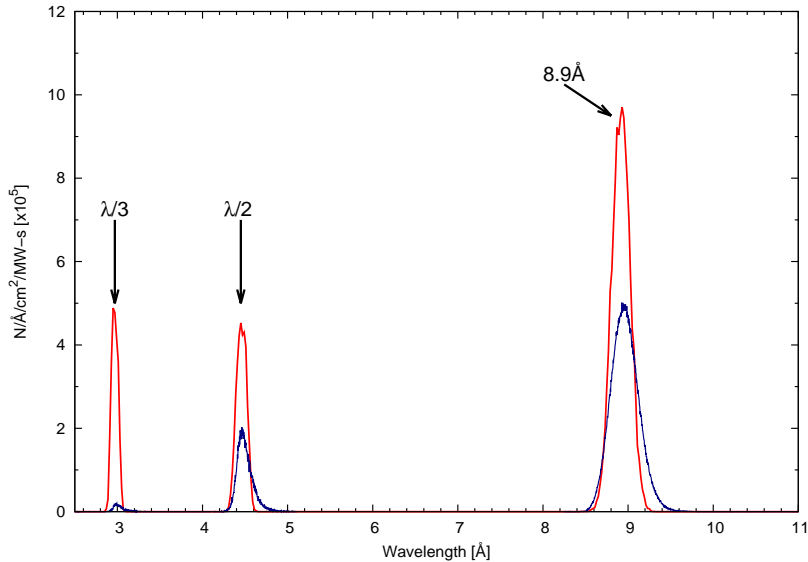}
\caption{Measured spectrum for the UCN guide at the FnPB.   A chopper
  upstream of the monochromator was used in order to separate neutrons
  of different wavelengths and several measurements at different chopper settings were combined to produce the final spectrum. Data were taken at a reduced repetition rate of 10 Hz.}
\label{ucn_spectrum}
\end{figure}

The spectrum of neutrons from the monochromatic beam is shown in Fig.~\ref{ucn_spectrum} along with the calculated flux from McStas.  In the model, the peak reflectivities of the first and second monochromators were taken to be 0.675 and 0.63, respectively.  In Ref.~\cite{Courtois2011S37}, the ranges of the crystal reflectivites are given as 70-80$\%$ for the potassium and 65-75$\%$ for the rubidium  monochromators. The middle of the range was used, with a 10$\%$ reduction due to neutron attenuation.  Additionally, in the Ref.~\cite{Courtois2011S37} measurement, each crystal was tested individually, whereas in the FnPB measurement, the crystals were arranged in 2D monochromator arrays.  To allow for possible misalignments, the peak reflectivities were scaled down by another 5$\%$.  The mosaics for each individual crystal were specified in the model.  The disagreement between the modelled and measured spectra at 8.9$\AA$ (the wavelength of interest) is almost a factor of two.  This is not currently understood.  The measurements reported in Ref.~\cite{Courtois2011S37} yield peak reflectivities that are down 15$\%$ (K-intercalated crystals) and 25$\%$ (Rb-intercalated crystals) from ideal values.  The reflectivities are a function of wavelength, possibly explaining a small fraction of the difference. It's also possible that some of the disagreement is due to the imperfect modeling of guide.   The decrease in measured flux from what was expected based on the McStas model has caused a modification in the planning of the nEDM experiment~\cite{edm_proposal,Golub:1994cg}, which is now expected to be using the cold beamline, BL13B.

\section{Summary}
The FnPB beamline at the Spallation Neutron Source has been comissioned and is now in operation for science experiments. Its measured performance is in reasonable agreement with simulations conducted in the design phase of the facility. 
Physics proposals are reviewed by the Fundamental Neutron Physics Proposal and Advisory Committee.  NPDGamma~\cite{npdg_proposal,snow00, Fomin:2013ssa}, the first of the approved peer-reviewed experiments, has recently been completed.  It will be followed by the $n-^3\mbox{He}$ hadronic parity violation experiment~\cite{nhe_proposal}, the Nab beta decay experiment~\cite{nab_proposal,dinko09}, and the nEDM experiment~\cite{edm_proposal,Golub:1994cg}.
\section{Acknowledgements}
W.~M.~Snow acknowledges support from NSF grants $\#$ NSF PHYS-0457219 and NSF PHYS-0758018, as well as the Indiana University Center for Spacetime Symmetries.  The material is based upon work supported by the U.S. Departmet of Energy, Office of Science, Office of Nuclear Physics, under award numbers DE-FG02-03ER41258 and DE-FG02-97ER41042.

\bibliographystyle{elsarticle-num-names}
\bibliography{beamline_main}

\begin{thebibliography}{34}
\providecommand{\natexlab}[1]{#1}
\providecommand{\url}[1]{\texttt{#1}}
\providecommand{\urlprefix}{URL }
\expandafter\ifx\csname urlstyle\endcsname\relax
  \providecommand{\doi}[1]{doi:\discretionary{}{}{}#1}\else
  \providecommand{\doi}[1]{doi:\discretionary{}{}{}\begingroup
  \urlstyle{rm}\url{#1}\endgroup}\fi
\providecommand{\bibinfo}[2]{#2}

\bibitem[{Nico and Snow(2005)}]{Nico05b}
\bibinfo{author}{J.~S. Nico}, \bibinfo{author}{W.~M. Snow},
  \bibinfo{title}{Experiments in Fundamental Neutron Physics},
  \bibinfo{journal}{Annual Reviews of Nuclear and Particle Science}
  \bibinfo{volume}{55} (\bibinfo{year}{2005}) \bibinfo{pages}{27}.

\bibitem[{Abele(2008)}]{Abele:2008}
\bibinfo{author}{H.~Abele}, \bibinfo{title}{The neutron. Its properties and
  basic interactions}, \bibinfo{journal}{Progress in Particle and Nuclear
  Physics} \bibinfo{volume}{60} (\bibinfo{year}{2008}) \bibinfo{pages}{1}.

\bibitem[{Dubbers and Schmidt(2011)}]{Dubbers:2011}
\bibinfo{author}{D.~Dubbers}, \bibinfo{author}{M.~G. Schmidt},
  \bibinfo{title}{The Neutron and Its Role in Cosmology and Particle Physics},
  \bibinfo{journal}{Reviews of Modern Physics} \bibinfo{volume}{83}
  (\bibinfo{year}{2011}) \bibinfo{pages}{1111}.

\bibitem[{Snow(2014)}]{Snow:2014}
\bibinfo{author}{W.~M. Snow}, \bibinfo{title}{Fundamental Neutron Physics with
  Long Pulsed Spallation Sources}, \bibinfo{journal}{Physics Procedia}
  \bibinfo{volume}{51} (\bibinfo{year}{2014}) \bibinfo{pages}{31}.

\bibitem[{Seo et~al.(2004)}]{Seo04}
\bibinfo{author}{P.~N. Seo}, et~al., \bibinfo{title}{A measurement of the
  Flight Path 12 cold H$_2$ moderator brightness at LANSCE},
  \bibinfo{journal}{Nuclear Instruments and Methods in Physics Research Section
  A} \bibinfo{volume}{517} (\bibinfo{year}{2004}) \bibinfo{pages}{285}.

\bibitem[{Arimoto et~al.(2012)Arimoto, Funahashi, Higashi, Hino, Hirota
  et~al.}]{Arimoto:2012zma}
\bibinfo{author}{Y.~Arimoto}, \bibinfo{author}{H.~Funahashi},
  \bibinfo{author}{N.~Higashi}, \bibinfo{author}{M.~Hino},
  \bibinfo{author}{K.~Hirota}, et~al., \bibinfo{title}{{Present status of
  neutron fundamental physics at J-PARC}},
  \bibinfo{journal}{Prog.~Theo.~Exp.~Phys.} \bibinfo{volume}{2012}
  (\bibinfo{year}{2012}) \bibinfo{pages}{02B007},
  \doi{\bibinfo{doi}{10.1093/ptep/pts075}}.

\bibitem[{Mishima et~al.(2009)}]{mishima09}
\bibinfo{author}{K.~Mishima}, et~al., \bibinfo{title}{Design of neutron
  beamline for fundamental physocs at J-PARC BL05},
  \bibinfo{journal}{Nucl.~Instr.~and Meth.~A} \bibinfo{volume}{600}
  (\bibinfo{year}{2009}) \bibinfo{pages}{342--345}.

\bibitem[{Fischer et~al.(1997)}]{Fischer97}
\bibinfo{author}{W.~E. Fischer}, et~al., \bibinfo{title}{SINQ - The spallation
  neutrons source, a new research facility at PSI}, \bibinfo{journal}{Physica
  B} \bibinfo{volume}{234} (\bibinfo{year}{1997}) \bibinfo{pages}{1202--1208}.

\bibitem[{Bauer(1998)}]{Bauer98}
\bibinfo{author}{G.~Bauer}, \bibinfo{title}{Operation and development of the
  new spallation neutron source SINQ at the Paul Sherrer Institute},
  \bibinfo{journal}{Nuclear Instruments and Methods in Physics Research Section
  B} \bibinfo{volume}{139} (\bibinfo{year}{1998}) \bibinfo{pages}{65}.

\bibitem[{Schebetov et~al.(2003)}]{Schebetov03}
\bibinfo{author}{A.~Schebetov}, et~al., \bibinfo{title}{New Facility for
  fundamental research in nuclear physics with polarized cold neutrons at PSI},
  \bibinfo{journal}{Nuclear Instruments and Methods in Physics Research Section
  A} \bibinfo{volume}{497} (\bibinfo{year}{2003}) \bibinfo{pages}{479}.

\bibitem[{Zejma et~al.(2005)}]{Zejma05}
\bibinfo{author}{J.~Zejma}, et~al., \bibinfo{title}{FUNSPIN polarized
  cold-neutron beam at PSI}, \bibinfo{journal}{Nuclear Instruments and Methods
  in Physics Research Section A} \bibinfo{volume}{539} (\bibinfo{year}{2005})
  \bibinfo{pages}{622}.

\bibitem[{Abele et~al.(2006)}]{Abele06}
\bibinfo{author}{H.~Abele}, et~al., \bibinfo{title}{Characterization of a
  ballisic supermirror neutron guide}, \bibinfo{journal}{Nuclear Instruments
  and Methods in Physics Research Section A} \bibinfo{volume}{562}
  (\bibinfo{year}{2006}) \bibinfo{pages}{407}.

\bibitem[{Nico et~al.(2005)}]{Nico05a}
\bibinfo{author}{J.~S. Nico}, et~al., \bibinfo{title}{The Fundamental Neutron
  Physics Facilities at NIST},
  \bibinfo{journal}{J.~Res.~Natl.~Inst.~Stand.~Technol} \bibinfo{volume}{110}
  (\bibinfo{year}{2005}) \bibinfo{pages}{137}.

\bibitem[{Cook(2009)}]{Cook09}
\bibinfo{author}{J.~C. Cook}, \bibinfo{title}{Design and estimated performance
  of a new neutron guide system for the NCNR expansion project},
  \bibinfo{journal}{J.~Res.~Natl.~Inst.~Stand.~Technol} \bibinfo{volume}{80}
  (\bibinfo{year}{2009}) \bibinfo{pages}{023101}.

\bibitem[{Stockli et~al.(2010)Stockli, Han, Murray, Pennisi, Santana, and
  Welton}]{stockli2010}
\bibinfo{author}{M.~P. Stockli}, \bibinfo{author}{B.~Han},
  \bibinfo{author}{S.~N. Murray}, \bibinfo{author}{T.~R. Pennisi},
  \bibinfo{author}{M.~Santana}, \bibinfo{author}{R.~F. Welton},
  \bibinfo{title}{Ramping up the Spallation Neutron Source beam power with the
  H- source using 0 mg Cs/day}, \bibinfo{journal}{Rev.~Sci.~Instrum.~}
  \bibinfo{volume}{81} (\bibinfo{year}{2010}) \bibinfo{pages}{02A729}.

\bibitem[{com(2006)}]{completion_report}
\bibinfo{title}{Spallation Neutron Source Project Completion Report},
  \bibinfo{year}{2006}.

\bibitem[{Henderson et~al.(2014)}]{Henderson2014}
\bibinfo{author}{S.~Henderson}, et~al., \bibinfo{title}{The Spallation Neutron
  Source accelerator system design}, \bibinfo{journal}{Nuclear Instruments and
  Methods in Physics Research Section A,} \bibinfo{note}{In Press}.

\bibitem[{Iverson et~al.(2002)Iverson, Ferguson, Gallmeier, and
  Popova}]{sns-sct-base02}
\bibinfo{author}{E.~B. Iverson}, \bibinfo{author}{P.~D. Ferguson},
  \bibinfo{author}{F.~X. Gallmeier}, \bibinfo{author}{I.~I. Popova},
  \bibinfo{title}{Detailed {SNS} neutronics calculations for scattering
  instrument design: {SCT} configuration}, \bibinfo{type}{Tech. Rep.}
  \bibinfo{number}{SNS 110040300-DA0001-R00}, \bibinfo{institution}{Oak Ridge
  National Laboratory}, \bibinfo{year}{2002}.

\bibitem[{Bowman(2005)}]{npdg_proposal}
\bibinfo{author}{J.~D. Bowman}, \bibinfo{title}{Precision Measurement of
  $A_{\gamma}$ in $n + p \rightarrow d + \gamma$}, \bibinfo{howpublished}{SNS
  proposal}, \urlprefix\url{http://npdgamma.com}, \bibinfo{year}{2005}.

\bibitem[{Snow et~al.(2000)}]{snow00}
\bibinfo{author}{W.~M. Snow}, et~al., \bibinfo{title}{Measurement of the parity
  Violating Asymmery in $n+p\rightarrow D+\gamma$},
  \bibinfo{journal}{Nucl.~Instr. and Meth.~A} \bibinfo{volume}{440}
  (\bibinfo{year}{2000}) \bibinfo{pages}{729--735}.

\bibitem[{Fomin(2013)}]{Fomin:2013ssa}
\bibinfo{author}{N.~Fomin}, \bibinfo{title}{{First results from the NPDGamma
  experiment at the spallation neutron source}}, \bibinfo{journal}{AIP
  Conf.Proc.} \bibinfo{volume}{1560} (\bibinfo{year}{2013})
  \bibinfo{pages}{145--148}, \doi{\bibinfo{doi}{10.1063/1.4826740}}.

\bibitem[{Lamoreaux and Golub(1995)}]{Lamoreaux:1995pq}
\bibinfo{author}{S.~Lamoreaux}, \bibinfo{author}{R.~Golub},
  \bibinfo{title}{{Recent progress in the development of a new technique to
  measure the neutron electric dipole moment}}  (\bibinfo{year}{1995})
  \bibinfo{pages}{597--604}.

\bibitem[{Golub and Pendlebury(1975)}]{golub1975}
\bibinfo{author}{R.~Golub}, \bibinfo{author}{J.~M. Pendlebury},
  \bibinfo{title}{Super-thermal sources of ultra-cold neutrons},
  \bibinfo{journal}{Phys.~Lett.~A}  (\bibinfo{year}{1975})
  \bibinfo{pages}{133--135}.

\bibitem[{Bowman and Po\v{c}ani\'{c}(2007)}]{nab_proposal}
\bibinfo{author}{J.~D. Bowman}, \bibinfo{author}{D.~Po\v{c}ani\'{c}},
  \bibinfo{title}{Precise Measurement of the Neutron Beta Decay Parameters, 'a'
  and 'b'}, \bibinfo{howpublished}{SNS proposal},
  \urlprefix\url{http://nab.phys.virginia.edu}, \bibinfo{year}{2007}.

\bibitem[{Po\v{c}ani\'{c} et~al.(2009)}]{dinko09}
\bibinfo{author}{D.~Po\v{c}ani\'{c}}, et~al., \bibinfo{title}{Nab: Measurement
  Principles, Apparatus and Uncertainties},
  \bibinfo{journal}{Nucl.~Instr.~Meth.~} \bibinfo{volume}{611}
  (\bibinfo{year}{2009}) \bibinfo{pages}{211--215}.

\bibitem[{Mattoni et~al.(2004)Mattoni, Adams, Alvine, Doyle, Dzhosyuk, Golub,
  Korobkina, McKinsey, Thompson, Yang, Zabel, and Huffman}]{Mattoni2004343}
\bibinfo{author}{C.~E.~H. Mattoni}, \bibinfo{author}{C.~P. Adams},
  \bibinfo{author}{K.~J. Alvine}, \bibinfo{author}{J.~M. Doyle},
  \bibinfo{author}{S.~N. Dzhosyuk}, \bibinfo{author}{R.~Golub},
  \bibinfo{author}{E.~Korobkina}, \bibinfo{author}{D.~N. McKinsey},
  \bibinfo{author}{A.~K. Thompson}, \bibinfo{author}{L.~Yang},
  \bibinfo{author}{H.~Zabel}, \bibinfo{author}{P.~R. Huffman},
  \bibinfo{title}{A long wavelength neutron monochromator for superthermal
  production of ultracold neutrons}, \bibinfo{journal}{Physica B: Condensed
  Matter} \bibinfo{volume}{344}~(\bibinfo{number}{1–4})
  (\bibinfo{year}{2004}) \bibinfo{pages}{343 -- 357}.

\bibitem[{Courtois et~al.(2011)Courtois, Menthonnex, Hehn, Andersen,
  Nesvizhevsky, Zimmer, Piegsa, Geltenbort, Greene, Allen, Huffman,
  Schmidt-Wellenburg, Fertl, and Mayer}]{Courtois2011S37}
\bibinfo{author}{P.~Courtois}, \bibinfo{author}{C.~Menthonnex},
  \bibinfo{author}{R.~Hehn}, \bibinfo{author}{K.~Andersen},
  \bibinfo{author}{V.~Nesvizhevsky}, \bibinfo{author}{O.~Zimmer},
  \bibinfo{author}{F.~Piegsa}, \bibinfo{author}{P.~Geltenbort},
  \bibinfo{author}{G.~Greene}, \bibinfo{author}{R.~Allen},
  \bibinfo{author}{P.~Huffman}, \bibinfo{author}{P.~Schmidt-Wellenburg},
  \bibinfo{author}{M.~Fertl}, \bibinfo{author}{S.~Mayer},
  \bibinfo{title}{Production and characterization of intercalated graphite
  crystals for cold neutron monochromators}, \bibinfo{journal}{Nuclear
  Instruments and Methods in Physics Research Section A}
  \bibinfo{volume}{634}~(\bibinfo{number}{1, Supplement})
  (\bibinfo{year}{2011}) \bibinfo{pages}{S37 -- S40}.

\bibitem[{Iverson et~al.(2005)Iverson, Micklich, Baxter, Cooper, Ferguson,
  Freeman, Gallmeier, Hammons, Lavelle, and Popova}]{ip:icansxvii-iverson-1}
\bibinfo{author}{E.~B. Iverson}, \bibinfo{author}{B.~J. Micklich},
  \bibinfo{author}{D.~V. Baxter}, \bibinfo{author}{R.~G. Cooper},
  \bibinfo{author}{P.~D. Ferguson}, \bibinfo{author}{D.~W. Freeman},
  \bibinfo{author}{F.~X. Gallmeier}, \bibinfo{author}{S.~E. Hammons},
  \bibinfo{author}{C.~M. Lavelle}, \bibinfo{author}{I.~Popova},
  \bibinfo{title}{Neutronic measurements to commission the {SNS}}, in:
  \bibinfo{booktitle}{Proceedings of {ICANS XVII}, the Seventeenth Meeting of
  the {International Collaboration on Advanced Neutron Sources}},
  \bibinfo{year}{2005}.

\bibitem[{Huffman et~al.(2005)Huffman, Greene, Allen, Cianciolo, Koehler,
  Desai, Mahurin, Yue, Palmquist, and Snow}]{huffman2005}
\bibinfo{author}{P.~R. Huffman}, \bibinfo{author}{L.~Greene, G.},
  \bibinfo{author}{R.~Allen}, \bibinfo{author}{V.~Cianciolo},
  \bibinfo{author}{P.~Koehler}, \bibinfo{author}{D.~Desai},
  \bibinfo{author}{R.~Mahurin}, \bibinfo{author}{A.~Yue},
  \bibinfo{author}{G.~R. Palmquist}, \bibinfo{author}{M.~Snow, W.},
  \bibinfo{title}{Beamline Performance Simulations for the Fundamental Neutron
  Physics Beamline at the Spallation Neutron Source}, \bibinfo{journal}{Journal
  of Research of the National Institute of Standards and Technology}
  \bibinfo{volume}{110} (\bibinfo{year}{2005}) \bibinfo{pages}{161--168}.

\bibitem[{Ito et~al.(2006)Ito, Crawford, and Greene}]{tito06}
\bibinfo{author}{T.~M. Ito}, \bibinfo{author}{C.~B. Crawford},
  \bibinfo{author}{G.~L. Greene}, \bibinfo{title}{Optimization of the Ballistic
  Guide Design for the SNS FNPB 8.9 A Neutron Line},
  \bibinfo{journal}{Nucl.~Instr.~Meth.~} \bibinfo{volume}{A564}
  (\bibinfo{year}{2006}) \bibinfo{pages}{414--423}.

\bibitem[{Hamilton and Iverson(2012)}]{ip-hamilton-2012}
\bibinfo{author}{A.~C. Hamilton}, \bibinfo{author}{E.~B. Iverson},
  \bibinfo{title}{Diagnostic Use of Neutron-Sensitive Image Plates at {ORNL}
  Neutron Facilities}, in: \bibinfo{booktitle}{Proceedings of the Tenth
  International Meeting on Nuclear Applications of Accelerators---AccApp'11},
  \bibinfo{pages}{235}, \bibinfo{year}{2012}.

\bibitem[{Cooper and Lamoreaux(2002)}]{edm_proposal}
\bibinfo{author}{M.~Cooper}, \bibinfo{author}{S.~Lamoreaux}, \bibinfo{title}{A
  New Search for the Neutron Electric Dipole Moment},
  \bibinfo{howpublished}{DOE proposal},
  \urlprefix\url{http://www.phy.ornl.gov/nedm/}, \bibinfo{year}{2002}.

\bibitem[{Golub and Lamoreaux(1994)}]{Golub:1994cg}
\bibinfo{author}{R.~Golub}, \bibinfo{author}{K.~Lamoreaux},
  \bibinfo{title}{{Neutron electric dipole moment, ultracold neutrons and
  polarized He-3}}, \bibinfo{journal}{Phys.Rept.} \bibinfo{volume}{237}
  (\bibinfo{year}{1994}) \bibinfo{pages}{1--62}.

\bibitem[{Bowman et~al.(2007)Bowman, Crawford, and Gericke}]{nhe_proposal}
\bibinfo{author}{J.~D. Bowman}, \bibinfo{author}{C.~Crawford},
  \bibinfo{author}{M.~Gericke}, \bibinfo{title}{A Measurement of the Parity
  Violating Proton Asymmetry in the Capture of Polarized Cold Neutrons on
  $^3$He}, \bibinfo{howpublished}{SNS proposal}, \bibinfo{year}{2007}.

\end{thebibliography}
\end{document}